\documentstyle[prl,aps]{revtex}
\begin{document}

\title{Measurement of the Deuteron Spin Structure Function 
g$_1^d(x)$ for 1~(GeV/c)$^2~<~Q^2~<~40$~(GeV/c)$^2$}

\author{
P.L.~Anthony,$^{16}$
R.G.~Arnold,${^1}$
T.~Averett,$^{5,*}$
H.R.~Band,$^{21}$
M.C.~Berisso,$^{12}$
H.~Borel,$^7$
P.E.~Bosted,${^1}$
S.L.~B${\ddot {\rm u}}$ltmann,$^{19}$
M.~Buenerd,$^{16,\dagger}$
T.~Chupp,$^{13}$
S.~Churchwell,$^{12,\ddagger}$
G.R.~Court,$^{10}$
D.~Crabb,$^{19}$
D.~Day,$^{19}$
P.~Decowski,$^{15}$
P.~DePietro,$^1$
R.~Erbacher,$^{16,17}$
R.~Erickson,$^{16}$
A.~Feltham,$^{19}$
H.~Fonvieille,$^3$
E.~Frlez,$^{19}$
R.~Gearhart,$^{16}$
V.~Ghazikhanian,$^6$
J.~Gomez,$^{18}$
K.A.~Griffioen,$^{20}$
C.~Harris,$^{19}$
M.A. Houlden,$^{10}$
E.W.~Hughes,$^5$
C.E.~Hyde-Wright,$^{14}$
G.~Igo,$^6$
S.~Incerti,$^3$
J.~Jensen,$^5$
J.R.~Johnson,$^{21}$
P.M.~King,$^{20}$
Yu.G.~Kolomensky,$^{5,12}$
S.E.~Kuhn,$^{14}$
R.~Lindgren,$^{19}$
R.M.~Lombard-Nelsen,$^7$
J.~Marroncle,$^7$
J.~McCarthy,$^{19}$
P.~McKee,$^{19}$
W.~Meyer,$^{4}$
G.~Mitchell,$^{21}$
J.~Mitchell,$^{18}$
M.~Olson,$^{9,\S}$
S.~Penttila,$^{11}$
G.~Peterson,$^{12}$
G.G.~Petratos,$^9$
R.~Pitthan,$^{16}$
D.~Pocanic,$^{19}$
R.~Prepost,$^{21}$
C.~Prescott,$^{16}$
L.M.~Qin,$^{14}$
B.A.~Raue,$^{8}$
D.~Reyna,$^{1}$
L.S.~Rochester,$^{16}$
S.~Rock,$^1$
O.A.~Rondon-Aramayo,$^{19}$
F.~Sabatie,$^7$
I.~Sick,$^2$
T.~Smith,$^{13}$
L.~Sorrell,$^1$
F.~Staley,$^7$
S.~St.Lorant,$^{16}$
L.M.~Stuart,$^{16,\parallel}$
Z.~Szalata,$^1$
Y.~Terrien,$^7$
A.~Tobias,$^{19}$
L.~Todor,$^{14}$
T.~Toole,$^1$
S.~Trentalange,$^{6}$
D.~Walz,$^{16}$
R.C.~Welsh,$^{13}$
F.R.~Wesselmann,$^{14}$
T.R.~Wright,$^{21}$
C.C.~Young,$^{16}$
M.~Zeier,$^2$
H.~Zhu,$^{19}$
B.~Zihlmann,$^{19}$}
\address{
{$^{1}$American University, Washington, D.C. 20016}  \break
{$^{2}$Institut f${\ddot u}$r Physik der Universit${\ddot a}$t Basel, 
CH-4056 Basel, Switzerland} \break
{$^{3}$University Blaise Pascal, LPC IN2P3/CNRS F-63170 Aubiere Cedex, France}
\break
{$^{4}$Ruhr-Universit${\ddot a}$t Bochum, Universit${\ddot a}$tstr. 150, 
Bochum, Germany} \break
{$^{5}$California Institute of Technology, Pasadena, California 91125}
\break
{$^{6}$University of California, Los Angeles, California 90095}
\break
{$^{7}$DAPNIA-Service de Physique Nucléaire, CEA-Saclay,
F-91191 Gif/Yvette Cedex, France} \break
{$^{8}$Florida International University, Miami, Florida 33199.} \break
{$^{9}$Kent State University, Kent, Ohio 44242} \break
{$^{10}$University of Liverpool, Liverpool L69 3BX, United Kingdom } \break
{$^{11}$Los Alamos National Laboratory, Los Alamos, New Mexico 87545} \break
{$^{12}$University of Massachusetts, Amherst, Massachusetts 01003} \break
{$^{13}$University of Michigan, Ann Arbor, Michigan 48109} \break
{$^{14}$Old Dominion University, Norfolk, Virginia 23529} \break
{$^{15}$Smith College, Northampton, Massachusetts 01063} \break
{$^{16}$Stanford Linear Accelerator Center, Stanford, California 94309 } \break
{$^{17}$Stanford University, Stanford, California 94305} \break
{$^{18}$Thomas Jefferson National Accelerator Facility, Newport News, Virginia
23606} \break
{$^{19}$University of Virginia, Charlottesville, Virginia 22901} \break
{$^{20}$The College of William and Mary , Williamsburg, Virginia 23187} \break
{$^{21}$University of Wisconsin, Madison, Wisconsin 53706} \break
}
\maketitle
\begin{abstract}
New measurements are reported on the deuteron spin structure function 
$g_{1}^{d}$.  These results were obtained from deep 
inelastic scattering of 48.3 GeV electrons on polarized deuterons in the 
kinematic range 0.01~$<$~$x$~$<$~0.9 and 
1~$<$~Q$^{2}$~$<$~40~(GeV/c)$^{2}$.  
These are the first high dose electron scattering data obtained using
lithium deuteride ($^{6}$Li$^{2}$H) as the target material.  
Extrapolations of the data were performed to obtain
moments of $g_{1}^{d}$, including $\Gamma_{1}^{d}$, and 
the net quark polarization $\Delta\Sigma$.
\end{abstract}
%
%
\twocolumn
Polarized deep inelastic scattering \cite{E143,E154,SMC,HERMES,E80etc}
is a powerful tool for 
studying the internal spin structure of the proton and neutron.  
The longitudinal spin structure
function
$g_{1}(x,Q^{2})$ so obtained
is sensitive to the quark spin distributions 
$\Delta$q($x,Q^{2}$)~=~q$^{\uparrow}$($x,Q^{2})~-~q^{\downarrow}$($x,Q^{2}$),
where the up (down) arrow refers to quark polarization parallel (anti-parallel)
to the nucleon polarization.  
These quantities depend on $x$, the fractional momentum carried by the struck 
parton,
and $Q^{2}$, the four-momentum transfer squared of the exchanged 
virtual photon.  
In this Letter, we report on new precision measurements 
of the deuteron 
spin structure function $g_{1}^{d}$
from the 
SLAC E155 experiment, which ran in early 1997.
These measurements,  covering
the range $ 0.01<x<0.9$ and $1<Q^{2}<40$ (GeV/c)$^{2}$,
are a significant 
improvement over previous deuteron measurements at SLAC \cite{E143}.
E155 is the first high dose electron scattering experiment to use polarized 
$^{6}$Li$^{2}$H (i.e.~$^{6}$LiD) as the deuteron 
target material.
These data on an isoscalar target provide a particularly strong 
constraint on the net 
quark polarization, $\Delta\Sigma~=~\Delta u+\Delta d+\Delta s$.
\par
The ratio of $g_{1}$ to the unpolarized structure function $F_{1}$
can be determined from measured longitudinal asymmetries $A_\parallel$
using
\begin{equation}
g_{1}/F_{1} = (A_{\parallel}/d) 
+ (g_{2}/ F_{1}) [ 2Mx/(2E - \nu)],
\label{EQ1}
\end{equation}
where $M$ is the nucleon mass, 
$d$~=~(1-$\epsilon$)(2-$y$)/[$y$(1+$\epsilon R$)], 
$y$~=~$\nu$/$E$~=~($E$-$E^{\prime}$)/$E$, $E$($E^{\prime}$) is the incident 
(scattered) electron energy,
$\epsilon^{-1}$~=~1~+~2[1+$\gamma^{-2}$]tan$^{2}$($\theta$/2),
$\theta$ is the electron scattering angle in 
the lab frame, $\gamma^{2}~=~Q^{2}$/$\nu^{2}$,
and $R(x,Q^{2})$ is the ratio of 
longitudinal to transverse 
virtual photon cross sections $\sigma_{L}/\sigma_{T}$. 
The longitudinal
asymmetry $A_{\parallel}$ is the cross section asymmetry 
between negative and
positive helicity electron beams incident on
target nucleons polarized parallel to the beam direction.  
In our analysis, 
the recent SLAC fit to $R$ \cite{R1998} and the NMC
fit to $F_{2}$ \cite{NMCF2} were used to calculate $F_{1}$. 
The contribution from the $g_{2}$ structure function was
determined using the $g_{2}^{WW}$ twist-two expression of Wandzura and 
Wilczek\cite{g2ww}, given by
\begin{equation}
g_{2}^{WW}(x,Q^{2}) = -g_{1}(x,Q^{2}) + \int_x^1 g_{1}(\xi,Q^{2})d\xi/\xi .
\end{equation}
This expression was initially evaluated 
using the 
E143 fit \cite{E143} for $g_{1}$, and the process was iterated including the
values for $g_{1}$ from E155.  Measurements of $g_{2}$ \cite{E155g2}
are consistent with $g_{2}^{WW}$.  The transverse
term in Eq.~\ref{EQ1} is suppressed  by the factor of 
$2Mx/(2E - \nu)$, so $g_{2}$ has only a few percent
impact on the $g_{1}$ results, even if $g_{2} = 0$ is used.
Although emphasis was given
to measurements with parallel target polarization to maximize the $g_{1}$
sensitivity, modest data samples with perpendicular 
target polarization were taken to measure $g_{2}$ \cite{E155g2}.
\par
Longitudinally polarized 48.3 GeV electron beam pulses 
of up to 400 ns duration 
were produced at 120~Hz by a 
circularly polarized laser beam illuminating a strained GaAs
photocathode.  The helicity for each pulse was selected by a 
32-bit pseudo-random number generator to minimize instrumental asymmetries.  
The beam polarization $P_{b}$ was determined periodically
during the data run using M{\o}ller scattering
from 20--154 $\mu$m thick Fe-Co-V polarized foils.
Results from independent single arm \cite{Moeller} and
coincidence detector systems agreed within 1\%.
No statistically significant deviations from the average value
of $P_{b}$ = 0.813~$\pm$~0.020 were seen during the experiment. \par

The $^{6}$LiD target \cite{Target} was 3 cm long, 
2.5 cm in diameter, and
enclosed in an aluminum cup.  
Lithium deuteride \cite{LiD}
provides a significant improvement over the 
previously used deuterated ammonia, because $^6$Li
can, to first order, be treated as a polarized deuteron plus an 
unpolarized alpha particle, and therefore half of the nucleons in $^6$LiD
are the desired polarizable species \cite{Rondon}.  It is also
fives times as radiation resistant as $^{15}$ND$_{3}$.
Pre-irradiation doses of 
1.3-4.5$\times$10$^{17}$e$^{-}$/cm$^{2}$ were used to create the
paramagnetic centers necessary for dynamic nuclear polarization.
At 1 K in a 5.004 T field, 
an average in-beam free deuteron
polarization $\langle P_{t} \rangle$ of 22\% was achieved using
140 GHz microwaves.
The maximum value was obtained after an exposure of approximately 
5$\times$10$^{15}$e$^{-}$/cm$^{2}$ in the 48 GeV beam.
The polarization of the target material was determined 
by nuclear magnetic resonance measurements
calibrated to the signal obtained at thermal equilibrium near
1.6 K.  An overall relative uncertainty of  4\% on P$_{t}$ was 
achieved using this technique.  
The $^{6}$Li polarization was measured to be 97\% of
the free deuteron polarization, consistent with predictions based on
equal spin temperatures and the ratio of magnetic
moments.  
Isotopic analysis of the target
material revealed that it contained molar fractions of 0.046 for $^{7}$Li 
and 0.024 for $^{1}$H.
Both of these impurities are polarizable and were accounted for in the
analysis. \par

Scattered electrons were detected in three independent magnetic spectrometers 
at central angles of 2.75$^\circ$, 5.5$^\circ$, and 10.5$^\circ$ with
respect to the incident beam.  
In the spectrometers at 2.75$^{\circ}$ and 5.5$^{\circ}$ \cite{E154}, 
electrons were identified using two threshold gas Cherenkov 
counters and an electromagnetic calorimeter consisting of 
a 20 by 10 stack of lead glass blocks 24 radiation lengths (X$_{0}$) thick.  
Particle momenta and scattering angles were measured 
with two sets of scintillator hodoscopes.  The 10.5$^\circ$  
spectrometer was added for E155 to double the $Q^{2}$ range of
the experiment.  It consisted of a single dipole between two 
quadrupoles followed by a single scintillator hodoscope, a threshold
gas Cherenkov counter, and an electromagnetic calorimeter with a preradiator
(PR)
section and a total absorption (TA) section.
The PR section consisted of 
ten 2X$_{0}$ thick
lead glass bars, which were also used in the particle momentum determination, 
while the TA section consisted of a 5 by 6 stack of lead glass blocks 
16X$_{0}$ thick.  
\par

The experimental longitudinal (A$_{\parallel}$)
asymmetries were determined
from the numbers of scattered electrons per incident beam charge
for negative ($N_{-}$) and positive ($N_{+}$) beam helicity using
\begin{equation}
A_{\parallel}^{d} =
\left(
 {{\displaystyle N_{-} - N_{+}} \over {\displaystyle N_{-} + N_{+}}} 
{{\displaystyle 1}\over {\displaystyle fP_{b}P_{t}C_{1}}} - C_{2} 
A_{\parallel}^{p} \right) 
{{\displaystyle 1} \over {\displaystyle f_{RC}}} + A_{RC} .
\end{equation}
The rates $N_{-}$ and $N_{+}$ were corrected for
contributions from charge symmetric background processes, which 
were measured by reversing the 
spectrometer polarity. This asymmetry is consistent with zero, 
leading to dilution corrections of 10\%-15\% at the lowest $x$ bin
in each spectrometer and decreasing rapidly at higher $x$. 
The rates $N_{-}$ and $N_{+}$ were also
corrected for
mis-identified hadrons, which were typically about 2\% or 
less of the electron
candidates.  Corrections for the rate dependence
of the electron detection and reconstruction efficiency were a few percent
for each spectrometer. 
\par
The dilution factor $f$ is the fraction of events
originating in the free deuterons.  It was determined from
the composition of the target materials, containers, and NMR coils
in the beam, which 
contained 18\% free deuterium, 53\% $^{6}$Li, 14\% $^{4}$He, 11\% Al, 3\%
O, and
1\% N by weight.  The $x$ dependence of the cross sections
for these nuclei caused $f$ to vary from 0.18 at low $x$ 
to 0.20 at high $x$.  
The relative systematic uncertainty on $f$ of $\sim$~2\% was dominated by
the uncertainties of the density and packing fraction of the 
$^{6}$LiD granules. 
\par
The factors $C_1$ and $C_2$ account for the
presence of several polarizable nuclear species in the target. 
$C_1$ includes the contributions from the free deuterons 
and $^{6}$Li, and ranges from 1.77 to 1.85, with 
an uncertainty of $\pm 0.05$.
$C_{1}f$ then gives an effective dilution factor of 
$\sim$~0.36, as compared with $\sim$~0.22 
for $^{15}$ND$_{3}$. 
Calculations based
on isospin conservation, shell model, solutions of Fadeev 
equations and a Green's function Monte Carlo provide consistent
values for the valence nucleon polarization \cite{Rondon}. 
From these analyses, we conclude that the effective deuteron in $^6$Li
has a net polarization of 87\% of the Li polarization.
The $C_2$ term
accounts for contributions from polarized protons in Li$^{1}$H and 
$^7$Li. This
correction is the product of the fraction of the
species present in the granules and a factor for the effective polarization of
the protons. The spin-3/2 $^{7}$Li nucleus
is described well in terms of two clusters: an $\alpha$ particle and a triton.
The neutrons in the triton are predominantly anti-aligned, so the
only significant polarized contribution comes from the proton \cite{Rondon}.
Values for $C_2$ range from $-0.023$ to $-0.030$ with an uncertainty
of $\pm 0.003$.  The proton asymmetry $A_{\parallel}^{p}$ was obtained
from a fit to world data.
\par
Both internal \cite{RCinternal} and external \cite{RCexternal}
radiative corrections were obtained using an iterative global fit of
all data \cite{E143,E154,SMC,HERMES,E80etc}, including E155.  Previous 
SLAC data were recorrected in a 
self-consistent manner.  The radiative multiplicative factor $f_{RC}$ and  
additive 
correction $A_{RC}$ were determined in a manner similar to that 
used in E154 \cite{E154}.  
For $x$ above $\sim 0.3$,
 A$_{\parallel}$ is changed by 
less than 2\% of its value, while at low $x$ it is decreased by
$\sim$ -0.005.  The $x$ and $Q^{2}$ dependent systematic uncertainties
on $A_{\parallel}$ due to the radiative corrections were typically 0.001,
0.002, and 0.004 for the 2.75$^{\circ}$, 5.5$^{\circ}$, and 10.5$^{\circ}$
spectrometers, respectively.
\par

\begin{figure}
\vspace*{5.1in}
\includegraphics{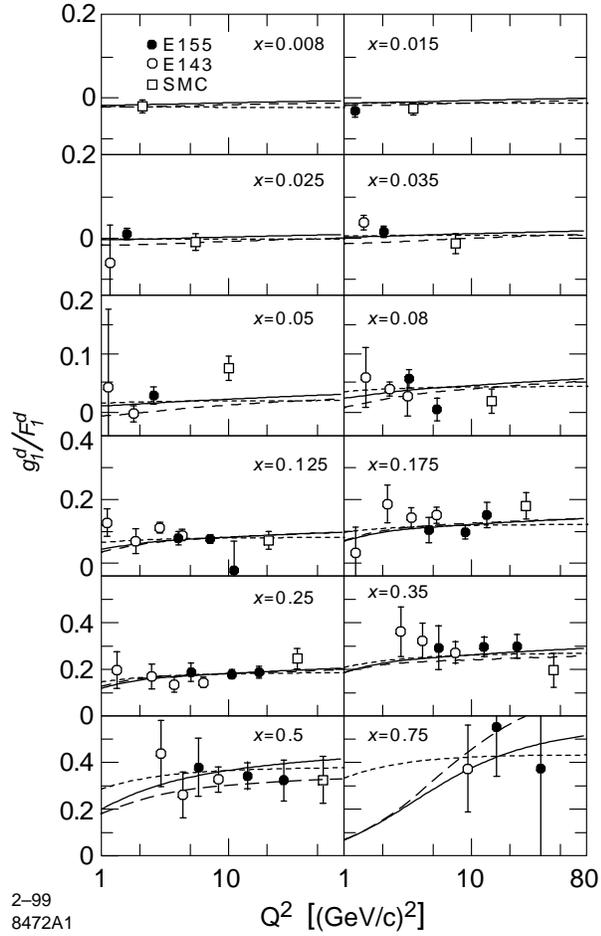}
\caption{
\label{fig:q2dep}
Values of $g_{1}^{d} / F_{1}^{d}$ vs. Q$^{2}$
at various values of $x$ for this experiment (filled circles),
 E143 (open circles), and SMC (open boxes). 
The inner bars are statistical
uncertainties, while the outer bars include systematic errors
added in quadrature.  
The curves are our parameterized fit (short-dashed), 
the E154 NLO fit \protect\cite{E154} (solid), and the
SMC NLO fit (long-dashed) \protect\cite{SMCNLO}.
}
\end{figure}

The E155 results for $g_{1}^{d}$/$F_{1}^{d}$ are shown in
Fig.~\ref{fig:q2dep} as a function of $Q^{2}$.
They are in good agreement with world data 
\cite{E143,E154,SMC,HERMES,E80etc} as demonstrated
by representative data from SMC and E143.  The data from the
three spectrometers provide both large $Q^{2}$ coverage and
good statistical resolution in the mid-$x$ region.
There is no significant $Q^{2}$ dependence
for $g_{1}^{d}/F_{1}^{d}$, indicating that the polarized and
unpolarized structure functions evolve similarly.  
\par
A simple parameterization was 
used to evolve the data to $Q^{2} = 5~({\rm GeV/c})^{2}$.
The world data from proton, neutron,
and deuteron targets \cite{E143,E154,SMC,HERMES,E80etc}, including E155,
with 
$Q^2>1$ (GeV/c)$^2$ and missing mass $W>2$ GeV were used to constrain
separate proton and neutron functions of the form
\begin{equation}
\label{EQ4}
(g_1 / F_1)=x^{\alpha}(a + bx + cx^2)(1+\beta/Q^2).
\end{equation}
The deuteron data were included in the fit using the relation
$g_{1}^{d} = {1 \over 2}
(g_{1}^{p} + g_{1}^{n})(1-1.5\omega_{D})$, where the deuteron 
D-state probability is $\omega_{D}$~=~0.05~$\pm$~0.01.
The variation \cite{Melnitchouk} of $\omega_{D}$ with $x$ is comparable 
to the uncertainty in $\omega_{D}$ for $x<0.75$ and negligible 
compared to the statistical uncertainties for $x>0.75$. 
The E155 deuteron 
data were included in 38 $x$ bins to preserve shape information and 
ensure proper treatment of 
the errors.  The parameters obtained
from the fit for the proton (neutron) were 
$\alpha~=~0.615~(-0.082)$, $a~=~0.715~(-0.056)$, $b~=~1.331~(-0.319)$,
$c~=~-1.766~(0.830)$, and $\beta~=~-0.17~(-0.14)$.
The $\beta$ values of $-0.17~\pm~0.05$ and $-0.14~\pm~0.30$
are small and consistent
with the data in Fig.~\ref{fig:q2dep}.   
The overall fit has a $\chi^2$ of 493 for 493 degrees of freedom.
\par
\begin{figure}
\vspace*{3.7in}
\includegraphics{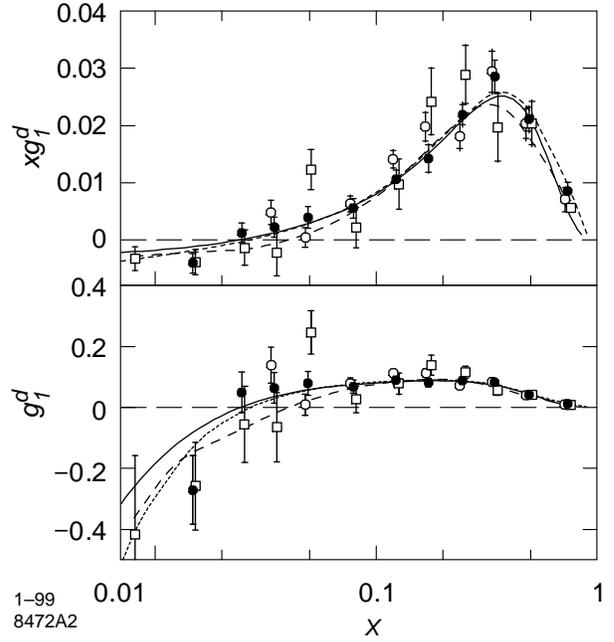}
\caption{
\label{fig:xdep}
Values of $xg_{1}^{d}$ (a) and $g_{1}^{d}$ (b) evolved to  
Q$^{2}$ of 5 (GeV/c)$^{2}$ using Eq. \protect\ref{EQ4}.  The
symbols and curves are as in Fig.~\protect\ref{fig:q2dep}.
}
\end{figure}

This fit and $F_{1}$
from \cite{R1998,NMCF2} were used to evolve the measured $g_{1}^{d}$ data
to a fixed 
$Q^2_0 = 5~({\rm GeV/c})^2$, shown in Fig.~\ref{fig:xdep} along 
with $xg_{1}^{d}$.
Integrating $g_{1}$ over the data range
yields $\int_{0.01}^{0.9} g_{1}^{d}(x,Q^2_0)dx$~
= $0.0401\pm0.0025({\rm stat})\pm0.0024({\rm syst})$.  
The extrapolation for the high $x$ 
contribution was found to be negligible.
However, the contribution from the low $x$ region from 0 to 0.01 
does not converge for Eq. \ref{EQ4}, reinforcing the need for 
additional data at very low $x$. 
Using the E154 \cite{E154} and SMC \cite{SMCNLO} perturbative QCD fits
 for the low $x$ extrapolation gives values of 
$\int_{0}^{1} g_{1}^{d}(x,Q^{2})~dx$~= 
$\Gamma_{1}^{d}$~=~$0.0266\pm~0.0025({\rm stat})\pm~0.0071({\rm syst})$ and 
$\Gamma_{1}^{d}$~=~$0.0291$ with comparable uncertainties, respectively.
\par
Using the expression obtained from the operator product expansion \cite{E143}, 
and information on the weak hyperon decay constants from \cite{PDG},
$\Delta\Sigma$ can be extracted
from 
$\Gamma_{1}^{d}$.  
Using the E154 and SMC fits for the low $x$ region, values of
$\Delta\Sigma = 0.15\pm0.03\pm0.08$ and $0.18$ 
(respectively)
are obtained in the $\overline{MS}$ scheme.  
These values are well below the 
Ellis-Jaffe prediction of 0.58 
\cite{EllisJaffe},
 but are in agreement with 
previous experimental values.  The precision
obtained using only the E155 deuteron data is comparable with results such as
the E154 NLO fit result of $0.020 \pm 0.06 \pm 0.05$ obtained
using all previous world data.  Improving the precision of $\Delta
\Sigma$ further will require a better understanding of the very low 
$x$ behavior
of $g_1$.
\par
The higher moments of $g_{1}$ are
also of theoretical interest and are less sensitive to the extrapolation
at low $x$.  Integrating $xg_{1}$ and $x^{2}g_{1}$ over $x$ at
$Q^2_0 = 5~({\rm GeV/c})^2$ yields
$0.0134~\pm~0.0007({\rm stat})~\pm~0.0008({\rm syst})$ and 
$0.0057~\pm~0.0004({\rm stat})~\pm~0.0004({\rm syst})$ in the data region. 
The contributions from the low and high $x$ regions are negligible.
Lattice QCD calculations \cite{Lattice} predict  
$\int_{0}^{1} x^{2}g_{1}^{d}(x,Q^{2} \sim 4~{\rm (GeV/c)}^{2})~dx~=~
0.0064~\pm~0.0017$,
which is in good agreement with $0.0061 \pm 0.0004 \pm 0.0004$
obtained by evolving the E155 data to $Q^{2} = $ 4~(GeV/c)$^{2}$.
\par
We sincerely thank the SLAC Experimental Facilities group for
their assistance in setting up the experiment and the SLAC Accelerator 
Operators group for their efficient beam delivery.  
This work was supported by the Department of Energy
(TJNAF, Massachusetts, ODU, SLAC, Stanford, Virginia, Wisconsin,
and William and Mary); the National Science
Foundation (American, Kent, Michigan, and ODU); the
Kent State University Research Council; the
Schweizersche Nationalfonds (Basel); 
the Commonwealth of Virginia (Virginia); 
the Centre National de la Recherche Scientifique and the 
Commissariat a l'Energie Atomique (French groups).

\end{document}